\documentclass[a4paper,oneside,11pt]{scrartcl}
\usepackage[T1]{fontenc}
\usepackage[ansinew]{inputenc}
\usepackage{lmodern}
\usepackage[a4paper,includehead,includefoot]{geometry}
\usepackage{natbib}
\usepackage{hyperref}
\usepackage{amsmath}
\usepackage{amsthm}
\usepackage{amssymb}
\usepackage{amsfonts}
\usepackage{bbm}
\usepackage{graphicx,booktabs}
\allowdisplaybreaks
\usepackage{setspace}
\usepackage{enumitem}

\begin{document}
\begin{flushleft}
\sffamily{\textbf{\LARGE{Shifting attention to old age: Detecting
  mortality deceleration using focused model selection}}} 
\vskip0.5cm
\normalfont
\normalsize

Marie B\"ohnstedt\footnotetext[1]{\textit{Address for correspondence:} M.~B\"ohnstedt, Max Planck Institute for Demographic Research, Konrad-Zuse-Str.~1, 18057 Rostock, Germany\\\sffamily{E-mail: boehnstedt@demogr.mpg.de}}\\
\small{\textit{Max Planck Institute for Demographic Research, Rostock, Germany\\Department of Biomedical Data Sciences, Leiden University Medical Center, Leiden, The Netherlands}}
\vskip0.25cm\normalsize
Hein Putter\\
\small{\textit{Department of Biomedical Data Sciences, Leiden University Medical Center, Leiden, The Netherlands}}\normalsize
\vskip0.25cm\normalsize
Nadine Ouellette\\
\small{\textit{Department of demography, Universit\'e de Montr\'eal, Montreal, Canada}}
\vskip0.25cm\normalsize
Gerda Claeskens\\
\small{\textit{ORSTAT and Leuven Statistics Research Center, KU Leuven, Leuven, Belgium}}
\vskip0.25cm\normalsize
Jutta Gampe\\
\small{\textit{Max Planck Institute for Demographic Research, Rostock, Germany}}\normalsize\\\vskip0.5cm\noindent
\textit{May 14, 2019}
\end{flushleft}\noindent
\textbf{Summary:} The decrease in the increase in death rates at old
ages is a phenomenon that has repeatedly been discussed in demographic research. While
mortality deceleration can be explained in the gamma-Gompertz model as
an effect of selection in heterogeneous populations, this
phenomenon can be difficult to assess statistically because it relates
to the tail of the distribution of the ages at death. By using a focused information
criterion (FIC) for model selection, we can directly target model
performance at those advanced ages. The gamma-Gompertz model is reduced
to the competing Gompertz model without mortality deceleration if the
variance parameter lies on the boundary of the parameter space. We
develop a new version of the FIC that is adapted to this non-standard
condition. In a simulation study, the new FIC is shown to outperform
other methods in detecting mortality deceleration.
The application of the FIC to extinct
French-Canadian birth cohorts demonstrates that focused model
selection can be used to rebut previous assertions about mortality deceleration.
\vskip0.25cm\noindent
\textit{Keywords:} Boundary; Focused information criterion; Gamma-Gompertz model; Heterogeneity; Model selection; Mortality deceleration

\section{Introduction}\label{sec:Intro}

The tendency for the death rates of human adults to increase exponentially with age is an
empirical regularity that has been known at least since
\citet{Gompertz:1825}. More recently, however, as vital
statistics have improved and more detailed information has become available for
individuals who survive to very old ages, a downward deviation from the exponential hazard  has been observed at advanced ages, and new mortality models for the oldest-old have been proposed
\citep{Thatcheretal:1998,Thatcher:1999,Bebbingtonetal:2014}.
This slowdown in the death rates at old ages is a phenomenon commonly known as
mortality deceleration.

Theoretically, such a deceleration is expected to occur if birth cohort members have heterogeneous mortality risks \citep{Beard:1959}. As a result of selection, frail individuals with higher mortality levels tend to die at younger ages, while the more robust individuals with lower death risks tend to survive to higher ages. This heterogeneity hypothesis has been supported empirically for human (and non-human) populations by a large number of studies \citep[among others][]{Vaupel:1979,HoriuchiWilmoth:1998,LynchBrown:2001}. However, the phenomenon of mortality deceleration has also been contested \citep{Gavrilov:2011,Gavrilova:2015}.

Accurately describing the age trajectory of mortality at
advanced ages is of interest to demographers and ageing researchers
because it provides insights into the
potential limits of human longevity \citep{Rootzen:2017}. However,
mortality deceleration also has
important implications for life insurance and public health.

\begin{figure}[hbt]
  \centering
 \includegraphics[width=\textwidth]{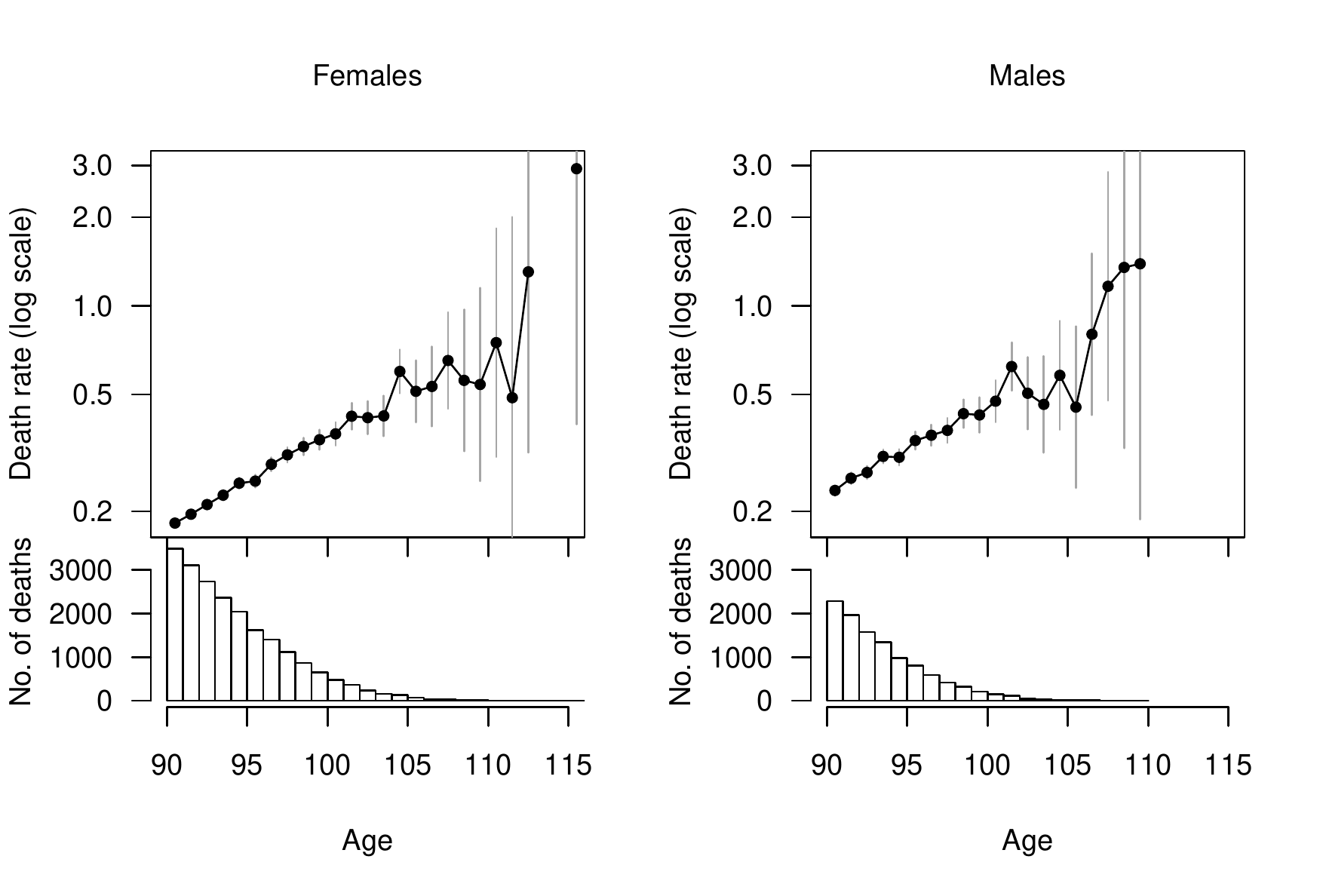}
\caption{Top: Death rates (on log scale) with 95\% confidence intervals for French-Canadian females (left) and males (right). Bottom:
  Frequency distribution of ages at death.}
\label{fig:Quebec1}
\end{figure}

As mortality deceleration occurs in the tail of the survival
distribution, where data are unavoidably scarce, the statistical
assessment of this phenomenon is challenging, and standard methods may fail to identify
deviations from  the Gompertz hazard for the very
old. For example, Figure~\ref{fig:Quebec1} shows for
French-Canadians born between 1880 and 1896 the empirical death rates by age at ages 90 and above,
as well as the number of deaths at each age. We can see that 75\% of
all deaths in this population had already occurred by age 96 for women and by age 95 for men.

Moreover, the apparent slowdown in death rates with advancing
age has repeatedly been attributed to data of questionable
quality. Exaggeration in the reporting of age and the failure to remove deceased
individuals from registers (due to unreported deaths) can result in an
overestimation of the number of long-lived individuals, which will bias death rates downward.
Thus, for individuals who die at very old ages, a thorough scientific
validation of the reported age at death is mandatory \citep{OdenseMonogr6}.
The need for age validation may, however, limit data availability. Here, we will analyse
a set of high-quality mortality data for French-Canadians 
born in Quebec at the end of the $19^{\text{th}}$ century. These data include
nominative information that is deemed confidential, but that is needed when carrying out
a rigorous age validation protocol. At the time of writing, we were granted access
to data only on individuals who had lived past the age of 90. Furthermore, adding the deaths of individuals aged 85-89 or 80-89 to the data set would have substantially increased the number of cases to be validated.
Such restrictions will make
the analysis considerably more demanding, as we will show. 

In this paper, we will discuss the statistical assessment of mortality
deceleration in the framework of the gamma-Gompertz model. This model
belongs to the class of proportional hazards frailty models, which
provide the standard approach for formalising the heterogeneity
hypothesis \citep{Vaupel:1979,Wienke:2011}. A Gompertz baseline hazard
is multiplied by a gamma-distributed random effect (the frailty).
The variance parameter of the gamma frailty describes the heterogeneity in the risk of death. If it takes a positive value, the individually heterogeneous hazards will result in a population hazard that shows a downward deviation from the exponentially increasing Gompertz hazard at advanced ages. If the variance parameter takes the value of zero, the population hazard is exponentially increasing; that is, the model is simplified to the Gompertz model. Thus, answering the question of whether mortality does or does not decelerate at advanced ages in this setting corresponds to selecting the gamma-Gompertz model or the Gompertz model. However, the single additional parameter of the gamma-Gompertz model -- namely, the variance parameter -- lies on the boundary of the parameter space if the true model is the Gompertz model. As this boundary constraint on the parameter violates the usual regularity assumptions, the inference and the model selection have to be adapted to this non-standard condition.

We propose using a focused information criterion
\citep[FIC,][]{ClaeskensHjort:2003} to assess mortality
deceleration. While other information criteria, like the Akaike
information criterion \citep[AIC,][]{Akaike:1974}, select a `best'
model regardless of the specific estimand that is of interest;
the FIC selects the model that performs `best' for a specific
parameter of interest, called the focus parameter. Applying the FIC
is particularly appealing in our application, as it will allow
us to choose a focus parameter that is directly affected by the
presence or absence of mortality deceleration; for example,
the hazard at some advanced age.
Technically, the FIC is constructed as an unbiased estimator of the
limiting risk of an estimator of the focus parameter, and the candidate
model with the smallest FIC value is selected. While the standard
version of the FIC aims to minimise the mean squared error (MSE) of
the estimator of the focus parameter, the criterion has been
generalised to other risk measures, such as $L_p$-risks
\citep{Claeskens:2006}. Still, all of these model selection criteria have been developed based on general likelihood theory under the standard regu\-lar\-ity assumptions, which are violated in our setting.
Therefore, we will derive versions of the FIC that allow us to choose between two
models in which the additional parameter may lie on the boundary of the
parameter space.

The paper is structured as follows. In Section~\ref{sec:MM}, we first
introduce the gamma-Gompertz model, and present traditional methods for
detecting mortality deceleration in this framework. Then, in Section~\ref{sec:FIC}, we propose the FIC as a new approach for assessing this phenomenon. In
Section~\ref{sec:simu}, we investigate the performance of the FIC in a
simulation study, and compare it with the performance of an AIC that is adjusted to the
presence of the boundary constraint. We apply the new model selection
criteria to the French-Canadian data in Section~\ref{sec:appl}, and we
conclude with a discussion in Section~\ref{sec:disc}.

\section{Mortality deceleration: Model and traditional approaches}\label{sec:MM}

\subsection{Gamma-Gompertz model}\label{subsec:Model}

To model adult lifespans (typically above age 30), we consider the
continuous random variable~$Y$. Its distribution can be characterised by the hazard function
\[h(y)=\lim_{\Delta y \searrow 0}=\frac{P(y<Y\leq y+\Delta y|Y>y)}{\Delta y}.\]
The standard approach to modelling individually heterogeneous hazards is via proportional hazards frailty models of the form~$h(y|Z=z)=z \cdot h_0(y)$. Here, a positive random effect~$Z$ (called frailty) acts multiplicatively on a common baseline hazard~$h_0(y)$, such that $h(y|Z=z)$ denotes the conditional hazard of an individual at age~$y$, given that his or her frailty is $Z=z$. The frailty~$Z$ is often assumed to follow a gamma distribution with mean one and variance~$\sigma^2$. The choice of the gamma distribution is both mathematically convenient and theoretically justified. \citet{AbbringvdBerg:2007} proved that the distribution of the heterogeneity among survivors converges to a gamma distribution for a large class of proportional hazards frailty models. The so-called gamma-Gompertz model is obtained if the gamma frailty is multiplied to an exponentially increasing Gompertz baseline hazard, $h_0(y)=a e^{by}$ with the parameters $a>0$ and $b>0$.

The variance parameter~$\sigma^2$ of the gamma-Gompertz model describes the heterogeneity of frailty. If $\sigma^2>0$, there is heterogeneity in the risk of death, and the selection of more robust individuals will take place. As a consequence, the resulting marginal hazard,
\begin{equation}\label{eq:GGhaz}
h(y)=\frac{a e^{by}}{1+\sigma^2 \frac{a}{b}(e^{by}-1)},
\end{equation}
shows a deceleration at advanced ages. If $\sigma^2=0$, there is no heterogeneity and the marginal hazard is exponentially increasing, such that $h(y)=a e^{by}$. Hence, in the framework of the gamma-Gompertz model, the statistical assessment of mortality deceleration is reduced to inference on the parameter~$\sigma^2$.

It is important to note that the parameter~$\sigma^2$ measures
population heterogeneity at the starting age of the model
(corresponding to $y=0$). Due to the
continuing selection of robust individuals, the variance of frailty among the survivors
decreases with age. Thus, the higher the age at which we start
our observation, the lower the heterogeneity in mortality is among the
individuals in the sample. The age at the beginning of the observation will,
therefore, have an impact on the resulting inference. 

The inference in the gamma-Gompertz model involves the frailty variance
$\sigma^2$, which is a parameter that lies on the boundary of its parameter
space in the absence of mortality deceleration ($\sigma^2=0$). This
violates the standard assumptions that underlie the asymptotic
properties of the likelihood-based inference, which in turn affects the
traditional approaches for assessing mortality deceleration that are
presented in the following section.

\subsection{Traditional approaches}\label{subsec:TradApp}

Two methods are commonly used for assessing mortality
deceleration in the framework of the gamma-Gompertz model: a likelihood ratio test for a zero frailty variance, and model selection between the gamma-Gompertz model and the Gompertz model based on the AIC.

The likelihood ratio test for homogeneity in the gamma-Gompertz model,
where $H_0\!: \sigma^2=0$ and $H_1\!: \sigma^2>0$, is non-standard
in that, under the null hypothesis, the parameter~$\sigma^2$ lies on
the boundary of the parameter space. Consequently, the asymptotic
distribution of the likelihood ratio test statistic under $H_0$ is no
longer a chi-squared distribution with one degree of
freedom. However, using the results of \citet{SelfLiang:1987}, it can
be shown that under the null hypothesis, the likelihood ratio test
statistic asymptotically follows a 50:50 mixture of a point mass at
zero and a chi-squared distribution with one degree of freedom,
$\frac{1}{2}\chi_0^2+\frac{1}{2}\chi_1^2$. Tests based on the wrong
assumption of a $\chi_1^2$-distribution of the test statistic
occasionally appear in studies of mortality decele\-ra\-tion
\citep{Pletcher:1999}.
Ignoring the issue of the boundary parameter and using the incorrect
distribution of the test statistic lowers the power to (correctly)
decide in favour of the gamma-Gompertz model.
But even when the test statistic is correctly assumed to follow a $\frac{1}{2}\chi_0^2+\frac{1}{2}\chi_1^2$-distribution, the likelihood ratio test has low power to detect mortality deceleration in the gamma-Gompertz model. This is especially likely to be the case when the inference has to be based on age-restricted samples, such as a sample of individuals who survived beyond age 90 (see Section~S.2 of the supplementary material for an illustration).

A popular alternative approach for assessing mortality deceleration is model
selection based on the AIC \citep{Richards:2008, Gavrilova:2015}. The
AIC targets an unbiased estimate of the Akaike information; that is,
of the expected relative Kullback-Leibler distance between the true
data-generating mechanism and the best parametric approximation. Under
standard conditions, the AIC is therefore defined as $-2\ell+2k$,
where the log-likelihood~$\ell$, evaluated at the maximum likelihood
estimate, is penalised by the number~$k$ of parameters in the
model. This common definition has, however, been found to be biased under the
non-standard conditions of the gamma-Gompertz model
\citep{Boehnstedt:2018}. Thus, the standard version of the AIC is not
a valid tool for model selection in the setting of the gamma-Gompertz
model. In Section~\ref{sec:AIC}, we will present a modified version of
the AIC that is adjusted to the presence of a boundary parameter.

\section{Focused information criterion for mortality deceleration}\label{sec:FIC}

The preceding considerations indicate that neither a testing
strategy, particularly if it is low-powered, nor an all-purpose model
selection criterion will adequately assess the occurrence of mortality
deceleration. Focused information criteria (FIC) have been
introduced to address problems of this kind, and we propose selecting
the model based on a new version of the FIC that takes the
boundary constraint on the frailty variance into account. 

\subsection{Rationale for FIC}
\label{sec:FICgen}

Statistical analyses are performed for particular purposes, and
acknowledging the specific purpose when choosing the statistical model
is the key concept of a FIC. In the following exposition, we use the
terminology and notation of \cite{ClaeskensHjort:2003}.

Observations $y_i, i=1,\ldots, n$ (here: ages at death) are assumed to
be generated by a parametric density $f(y)$. The parameters of the
model are split into a $d$-vector $\boldsymbol{\theta}$, which characterises the
narrow model, and an additional $q$-vector $\boldsymbol{\gamma}$ for the
extended model. The narrow model is obtained for one particular value
$\boldsymbol{\gamma_0}$, which is fixed and known. In the current application, the density of the gamma-Gompertz
model~\eqref{eq:GGhaz} is
\[f(y)=a
  e^{by}\left\{1+\sigma^2\frac{a}{b}(e^{by}-1)\right\}^{-\left(1+\frac{1}{\sigma^2}\right)}.\]
The parameter $\boldsymbol{\theta}=(a,b)^T$ is the Gompertz part of
the model, so $d=2$. The single additional parameter is
$\gamma=\sigma^2$ with $\gamma_0=0$, so $q=1$.

The original FIC is derived in a framework of local misspecification \citep{HjortClaeskens:2003}, where a sample of size~$n$ is assumed to be generated from a density
\begin{equation}\label{eq:misspec}
f_{\text{true}}(y)=f(y,\boldsymbol{\theta}_0,\boldsymbol{\gamma}_0+\boldsymbol{\delta}/\sqrt{n}),
\end{equation}
with the parameter
vector~$\boldsymbol{\gamma}=\boldsymbol{\gamma}_0+\boldsymbol{\delta}/\sqrt{n}$
perturbed in the direction of~$\boldsymbol{\delta}$. Selection is
between the null model, where $\boldsymbol{\gamma}$ is fixed at the
known value $\boldsymbol{\gamma}_0$; the full model, including both
$\boldsymbol{\theta}$ and $\boldsymbol{\gamma}$; and, if $q>1$, any
model including~$\boldsymbol{\theta}$, but only a subset of the
components of $\boldsymbol{\gamma}$ and the remaining fixed at the
respective values in $\boldsymbol{\gamma}_0$.
For the current setting, selection is only between the null model with $\sigma^2=0$, that is, the Gompertz model; and the full model including $\sigma^2$, that is, the gamma-Gompertz model.
Due to the boundary constraint on the frailty variance, $\delta=\sqrt{n}\sigma^2$ is subject to the a priori restriction $\delta\geq 0$. Therefore, we will restrict the framework in the following to the choice of including or not including a single parameter with a boundary constraint; that is, $q=1$ and $\gamma\geq \gamma_0$.

The focus is the parameter of interest, which depends on the
underlying density~\eqref{eq:misspec} via $\boldsymbol{\theta}$ and
$\boldsymbol{\gamma}$. The focus is commonly denoted by $\mu$, and we define
$\mu_{\text{true}}=\mu(\boldsymbol{\theta},\gamma_0+\delta/\sqrt{n})$.
Based on the maximum likelihood estimators
$\boldsymbol{\hat{\theta}}_{\text{null}}$ in the null model and
$(\boldsymbol{\hat{\theta}}_{\text{full}},\hat{\gamma})$ in the full
model, the focus parameter is estimated as
$\hat{\mu}_{\text{null}}=\mu(\boldsymbol{\hat{\theta}}_{\text{null}},\gamma_0)$
or
$\hat{\mu}_{\text{full}}=\mu(\boldsymbol{\hat{\theta}}_{\text{full}},\hat{\gamma})$.
For each model~$M$, $M\in\{\text{null, full}\}$, the
estimator~$\hat{\mu}_M$ converges in distribution,
$\sqrt{n}(\hat{\mu}_M-\mu_{\text{true}})\stackrel{d}{\longrightarrow}\Lambda_M$.

The FIC selects the model that performs `best' for the focus
parameter~$\mu$. If it is based on the general $L_p$-loss, the FIC aims to
estimate without bias the limiting $L_p$-risk of $\hat{\mu}_M$; that is,
$r_p(M)=\mathbb{E}[|\Lambda_M|^p]$. The model for which
this limiting risk is smaller is selected by the criterion.
Of particular interest is a FIC based on the MSE \citep[$p=2$, as for
the original version,][]{ClaeskensHjort:2003}, constructed as an
estimator of $\mathbb{E}[\Lambda_M^2]$; and a FIC based on the mean
absolute error (MAE, $p=1$), constructed as an estimator of $\mathbb{E}[|\Lambda_M|]$.

\subsection{FIC with a parameter on the boundary of the parameter
  space}
\label{sec:FICboundary}

Under standard regularity conditions, when general likelihood theory applies, the asymptotic normality of the maximum likelihood estimator implies that the $\Lambda_M$ are normally distributed \citep{ClaeskensHjort:2003}. In the non-standard setting considered here, $\Lambda_{\text{full}}$ is not normally distributed because the maximum likelihood estimator $(\boldsymbol{\hat{\theta}}_{\text{full}},\hat{\gamma})$ converges in distribution to a mixture with two components \citep{Boehnstedt:2018}. The limiting distribution depends on the information matrix~$J_{\text{full}}$ of the full model evaluated at the null model~$(\boldsymbol{\theta}_0,\gamma_0)$. We denote by $J_{00}$, $J_{01}$, $J_{10}$, and $J_{11}$, the four blocks of $J_{\text{full}}$ corresponding to the components~$\boldsymbol{\theta}$ and $\gamma$ of the parameter vector; and by $\kappa^2$ the element of the inverse information matrix~$J_{\text{full}}^{-1}$, which corresponds to $\gamma$.
Then, the following convergence in distribution holds for the estimator of the frailty variance
\[\sqrt{n}(\hat{\gamma}-\gamma_0)\stackrel{d}{\longrightarrow} \max{(0,D)}\quad \text{with}~D\sim\mathcal{N}(\delta,\kappa^2).\]
For the limiting distribution of the estimator of the focus parameter, it can be shown that
\begin{align}\label{eq:lambdaS}
\sqrt{n}(\hat{\mu}_{\text{null}}-\mu_{\text{true}})&\stackrel{d}{\longrightarrow}\Lambda_{\text{null}}=\Lambda_0+\omega\delta\quad\text{and}\notag\\
\sqrt{n}(\hat{\mu}_{\text{full}}-\mu_{\text{true}})&\stackrel{d}{\longrightarrow}\Lambda_{\text{full}}=\begin{cases}\Lambda_0+\omega(\delta-D)&\text{if}~~D>0\\\Lambda_0+\omega\delta&\text{if}~~D\leq0\end{cases},
\end{align}
where $\Lambda_0\sim\mathcal{N}(0,\tau_0^2)$ is independent of $D$, $\tau_0^2=\left(\frac{\partial\mu}{\partial\theta}\right)^T J_{00}^{-1}\frac{\partial\mu}{\partial\theta}$ and $\omega=J_{10}J_{00}^{-1}\frac{\partial\mu}{\partial\theta}-\frac{\partial\mu}{\partial\gamma}$ \citep[cf.~Section~10.2 in][]{ClaeskensHjort:2008b}.

To define a FIC, we need to derive  $\mathbb{E}[|\Lambda|]$ or
$\mathbb{E}[\Lambda^2]$ from~\eqref{eq:lambdaS}, depending on whether
we intend to base the criterion on the limiting $L_1$- or $L_2$-risk
of the estimator $\hat{\mu}$.

As in the original version of the FIC, the limiting MSE of
$\hat{\mu}$ is considered first. However, as we will demonstrate in the
following, the FIC based on the $L_2$-risk has some drawbacks in the
current setting, which makes the $L_1$-risk an attractive alternative.

From equation~\eqref{eq:lambdaS} we can determine
$\mathbb{E}[\Lambda^2]$ for the null and the full model:
\begin{equation}\label{eq:MSE}
\mathbb{E}[\Lambda_{\text{null}}^2]=\tau_0^2+\omega^2\delta^2\quad\text{and}\quad\mathbb{E}[\Lambda_{\text{full}}^2]=\tau_0^2+\omega^2\left\{\delta^2\Phi\left(-\frac{\delta}{\kappa}\right)-\kappa\delta\phi\left(\frac{\delta}{\kappa}\right)+\kappa^2\Phi\left(\frac{\delta}{\kappa}\right)\right\},
\end{equation}
where $\Phi(\cdot)$ and $\phi(\cdot)$ denote the cdf and the pdf of the
standard normal distribution, respectively.
The $\text{FIC}_{\text{MSE}}$ would be constructed as an unbiased
estimator of the MSEs in~\eqref{eq:MSE}, and the model with the
smaller FIC value would be selected.

As has already been pointed out by \citet[Section
5.3]{ClaeskensHjort:2008b}, in the case of a
single additional parameter $\gamma$, the so-called tolerance radius
does not depend on the focus $\mu$. This radius signifies the
deviation $\delta$ for which the MSE of the null
model estimator is smaller than that of the full model estimator; that
is, $\mathbb{E}[\Lambda_{\text{null}}^2]\leq
\mathbb{E}[\Lambda_{\text{full}}^2]$. From~\eqref{eq:MSE}, we see that
the two risks are the same for $\omega=0$, and that if $\omega\neq 0$ the tolerance radius encompasses all $\delta$ with $\delta<0.8399\kappa$.
We can still define a pre-test strategy for assessing mortality
deceleration, which is based on the
quantity~$\hat{\delta}/\hat{\kappa}$, where
$\hat{\delta}=\sqrt{n}(\hat{\gamma}-\gamma_0)=\sqrt{n}\hat{\sigma}^2$
and $\hat{\kappa}$ is derived from the observed Fisher information. If
$\hat{\delta}/\hat{\kappa}\leq 0.8399$, the
estimator~$\hat{\mu}_{\text{null}}$ based on the Gompertz model is
used; whereas if $\hat{\delta}/\hat{\kappa}>0.8399$, the
estimator~$\hat{\mu}_{\text{full}}$ based on the gamma-Gompertz model
is used. We note here that $\hat{\delta}$ is not an unbiased
estimator of $\delta$, with the bias depending in a complex way on
$\delta$ and $\kappa$.
In appraising this pre-test-based model choice,
we can see that for large
samples, the local power of this strategy is approximately the same as the power of a likelihood ratio test
for $H_0\!: \sigma^2=0$ at the 20\% level (cf.~Section S.3 of the supplementary material).

Although strategies based on the limiting $L_2$-risks of the estimator $\hat\mu
$ are common, the derived pre-test strategy has drawbacks. On the one
hand, the performance of this strategy does not depend on the chosen focus parameter; while on the other, the equal penalty for squared bias and variance of the estimators in the $L_2$-risk might not be suitable for choosing whether to include a heterogeneity parameter.

Consequently, using risk measures other than the $L_2$-risk can be more
appropriate, as was already suggested in
\citet{Claeskens:2006}. Formulas for the general limiting $L_p$-risk
of $\hat{\mu}_M$ were derived there under regularity conditions where
$\Lambda_M$ follows a normal distribution for each of the models. In
our non-standard setting, the limiting distribution of the full model
estimator in~\eqref{eq:lambdaS} is not normal, but we can still derive
the limiting $L_1$-risk of the estimators $\hat{\mu}_{\text{null}}$
and $\hat{\mu}_{\text{full}}$ as follows (see Section S.4 of the supplementary material for details):
\begin{alignat}{2}
&\mathbb{E}[|\Lambda_{\text{null}}|]&&=2\tau_0\phi\left(\frac{\omega\delta}{\tau_0}\right) +2\omega\delta \left\{\Phi\left(\frac{\omega\delta}{\tau_0}\right)-\frac{1}{2}\right\}\quad\text{and}\notag\\
&\mathbb{E}[|\Lambda_{\text{full}}|]&&=\left[2\tau_0\phi\left(\frac{\omega\delta}{\tau_0}\right) +2\omega\delta \left\{\Phi\left(\frac{\omega\delta}{\tau_0}\right)-\frac{1}{2}\right\}\right]\left\{1-\Phi\left(\frac{\delta}{\kappa}\right)\right\}\label{eq:L1}\\
& &&\hspace{0.5cm}+\sqrt{\tau_0^2+\omega^2\kappa^2}\cdot
\sqrt{\frac{2}{\pi}}\Phi\left(\frac{\delta}{\kappa}\cdot
  \frac{\sqrt{\tau_0^2+\omega^2\kappa^2}}{\tau_0}\right)-\omega\kappa\,\phi\left(\frac{\delta}{\kappa}\right)
\cdot 2
\left\{\Phi\left(\frac{\omega\delta}{\tau_0}\right)-\frac{1}{2}\right\}\notag .
\end{alignat}
Thus, we define the $\text{FIC}_{\text{MAE}}$ of the null model and the full model as the estimators
\begin{alignat*}{2}
&\text{FIC}_{\text{MAE}}(\text{null})&&=2\hat{\tau}_0\phi\left(\frac{\hat{\omega}\hat{\delta}}{\hat{\tau}_0}\right) +2\hat{\omega}\hat{\delta} \left\{\Phi\left(\frac{\hat{\omega}\hat{\delta}}{\hat{\tau}_0}\right)-\frac{1}{2}\right\}~~~\text{and}\\
&\text{FIC}_{\text{MAE}}(\text{full})&&=\left[2\hat{\tau}_0\phi\left(\frac{\hat{\omega}\hat{\delta}}{\hat{\tau}_0}\right) +2\hat{\omega}\hat{\delta} \left\{\Phi\left(\frac{\hat{\omega}\hat{\delta}}{\hat{\tau}_0}\right)-\frac{1}{2}\right\}\right]\left\{1-\Phi\left(\frac{\hat{\delta}}{\hat{\kappa}}\right)\right\}\notag\\
& &&\hspace{0.5cm}+\sqrt{\hat{\tau}_0^2+\hat{\omega}^2\hat{\kappa}^2}\cdot \sqrt{\frac{2}{\pi}}\Phi\left(\frac{\hat{\delta}}{\hat{\kappa}}\cdot \frac{\sqrt{\hat{\tau}_0^2+\hat{\omega}^2\hat{\kappa}^2}}{\hat{\tau}_0}\right)-\hat{\omega}\hat{\kappa}\,\phi\left(\frac{\hat{\delta}}{\hat{\kappa}}\right) \cdot 2 \left\{\Phi\left(\frac{\hat{\omega}\hat{\delta}}{\hat{\tau}_0}\right)-\frac{1}{2}\right\},
\end{alignat*}
respectively.
Based on this new model selection criterion~$\text{FIC}_{\text{MAE}}$,
the full model is chosen if the estimated MAE of its estimator of the
focus parameter $\mu$ is smaller than the MAE for the null model estimator. In contrast to the MSE, the tolerance radius determined by the MAE of $\hat{\mu}_M$ does depend on the focus parameter via $\omega$ and $\tau_0$.

\subsection{Choice of the focus parameter}
\label{sec:FICfocus}
The central concept and virtue of the FIC approach is that it allows us
to consolidate a scientific question in a focus parameter, and to
customise the model selection to the specific focus. In the context of
mortality deceleration, two focus parameters suggest themselves. The first parameter is the
frailty variance, since it determines whether mortality deceleration is present, so  $\mu=\sigma^2$. The second focus parameter targets the deceleration of the hazard function, measured by the
second derivative of the log-hazard at some (high) age $y$ so that
$\mu =[ \ln h(y) ] '' $.

For $\mu = \sigma ^2$ the expressions in~\eqref{eq:L1} take the form
\[\mathbb{E}[|\Lambda_{\text{null}}|]=\delta\quad\text{and}\quad\mathbb{E}[|\Lambda_{\text{full}}|]=\kappa
  \sqrt{\frac{2}{\pi}}-\kappa\phi\left(-\frac{\delta}{\kappa}\right)+\delta\Phi\left(-\frac{\delta}{\kappa}\right).\]
Consequently, model choice based on the $\text{FIC}_{\text{MAE}}$
results in the gamma-Gompertz model if
$\hat{\delta}/\hat{\kappa}>0.6399$. If we view this as a pre-test
strategy, then it has asymptotically the same local power as the
likelihood ratio test for $H_0\!: \sigma^2=0$ at a level of 26\%.

If we choose $\mu =[ \ln h(y) ] '' $ the choice of the age $y$ should
be such that it marks an age in the tail of the distribution where
deceleration occurs, but which still lies within the range of observed lifespans.

While the above choices of the focus parameter are natural and allow for
immediate interpretations, we could also select as the focus any function
that characterises the distribution of lifespans, such as the survival
function or the log-hazard. The effects of different focus parameters on the model selection will be briefly illustrated in the simulation study in Section~\ref{sec:simu}, and recommendations will be given in Section~\ref{sec:disc}.

\subsection{A modified AIC for the gamma-Gompertz model}\label{sec:AIC}

As we mentioned in Section~\ref{subsec:TradApp}, the standard AIC is
biased as an estimator of the Akaike information in the presence of a
boundary parameter, and should therefore
not be used for assessing mortality deceleration. However,
\citet{Boehnstedt:2018} explicitly derived the bias of the standard AIC
for the gamma-Gompertz model~\eqref{eq:GGhaz} under the local
misspecification framework~\eqref{eq:misspec} as
$2\,\Phi\left(-\delta/\kappa\right)$. This bias depends via
$\delta=\sqrt{n}\sigma^2$ on the unknown value of the frailty
variance, and it cannot be estimated without bias if the true variance
is small. Thus, the bias cannot be removed completely, but it can be
reduced if we correct the standard AIC using the estimator $2\,\Phi\left(-\hat{\delta}/\hat{\kappa}\right)$ of the bias term. Hence, we define a modified version of the AIC for the gamma-Gompertz model as
\begin{equation}\label{eq:AIC*}
\text{AIC}^{\ast}=-2\,\ell+2\cdot 3-2\,\Phi\left(-\frac{\hat{\delta}}{\hat{\kappa}}\right).
\end{equation}
The performance of this modified $ \text{AIC}^{\ast}$ for detecting mortality deceleration is studied in the next section.

\section{Simulation study}\label{sec:simu}

To examine the performance of the proposed $\text{FIC}_{\text{MAE}}$
in assessing mortality deceleration, we conducted a simulation
study. In addition to considering different choices for the focus,
the study compares the behaviour of the $\text{FIC}_{\text{MAE}}$ with that of
the pre-test based on $L_2$-risks, and with that of the $\text{AIC}^{\ast}$
defined in~\eqref{eq:AIC*}.

The following factors will affect the performance of the
different strategies: the size of the true frailty variance
$\sigma^2$; the sample size $n$; and the starting age used when observing
lifespans, with a younger starting age being more
favourable for detecting actual mortality deceleration.

For the frailty variance (at $y=0$), three different scenarios were considered:
$\sigma^2=0.0625$ ($S_1$) and $\sigma^2=0.03$ ($S_2$) with Gompertz
parameters  $a=0.013$, $b=0.092$. Scenario~$S_3$ is a pure Gompertz
model with $a=0.0198$, $b=0.0726$ (and $\sigma^2=0$).
These numbers were inspired by the data on French-Canadian females that are analysed in the following section.

To cover the latter two aspects, survival times were generated from the
gamma-Gompertz model~\eqref{eq:GGhaz}, with $y=0$ corresponding to age
60. However, the model selection was based only on subsets of
individuals reaching certain ages. Motivated by
the French-Canadian data, we considered individuals who survived to ages 90 or higher
(90+). Additional comparisons based on the larger subsets
of individuals who survived to ages 85+ and 80+
are presented in Section S.6 of the supplementary material.

For each scenario~$S_1$ to $S_3$, three different initial
sample sizes (at age 60) were chosen, such that the size of the 90+
subset approximately equals $n_{90+}=$ 10,000 (small), $n_{90+}=$ 20,000
(medium) or $n_{90+}=$ 105,000 (large). The sample sizes may look
unusually large, but they cover a realistic range of population-based
data. The French-Canadian data presented in Figure~\ref{fig:Quebec1}
contain information on about 20,000 women and 10,000 men.

For each 90+ sample, the
log-likelihoods for the Gompertz model and the gamma-Gompertz model
were maximised using function \texttt{nlm()} in R \citep{R}; further
computational details are given in Section S.1 of the supplementary material. Then,
the best model is selected based on the $\text{FIC}_{\text{MAE}}$ for
different focus parameters, the MSE pre-test of
$\delta<0.8399\kappa$, and the $\text{AIC}^{\ast}$. We ran 1,000
replications for each setting.

\begin{figure}[hbt]
\centering
\includegraphics[width=\textwidth]{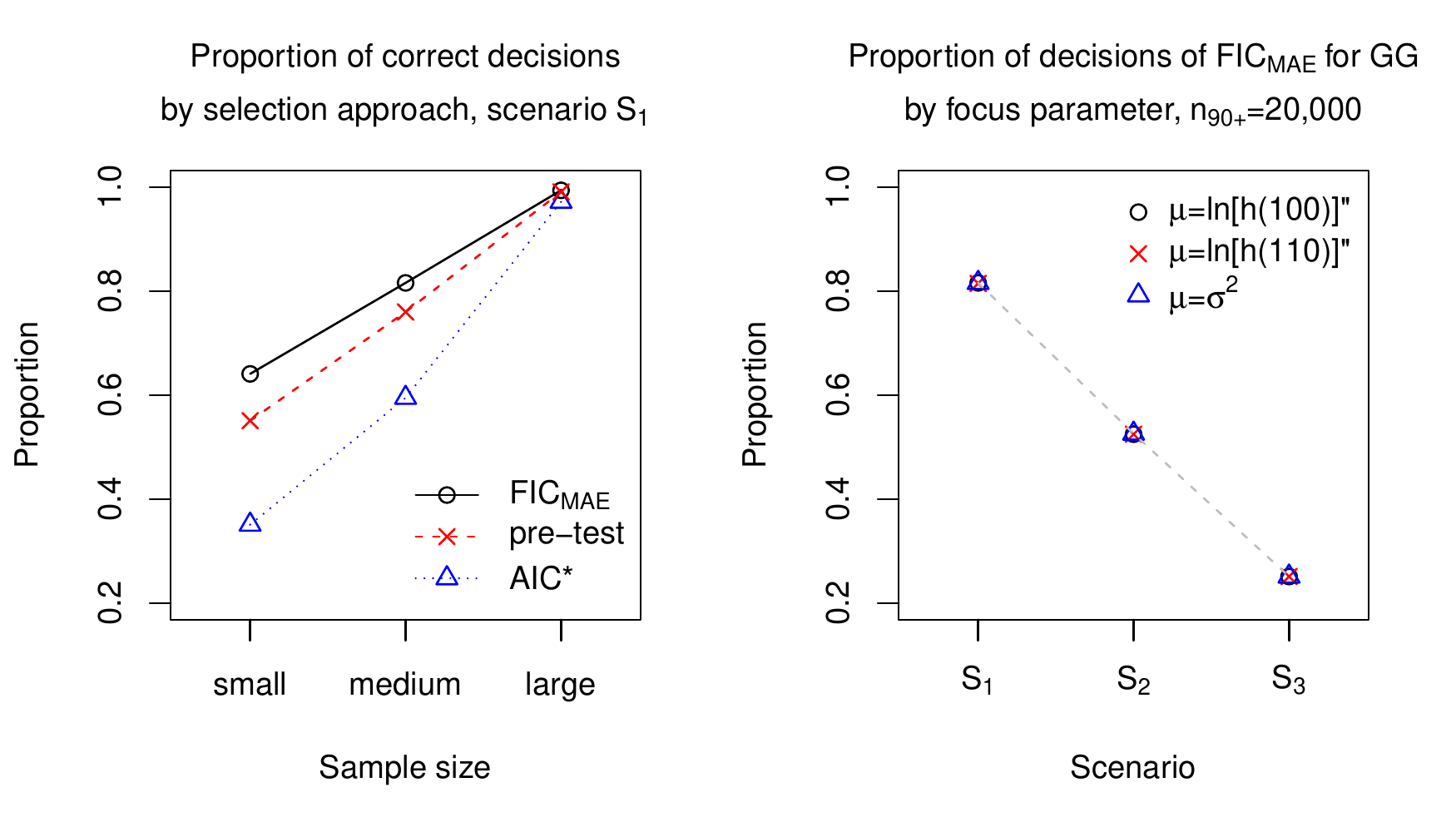}
\caption{Proportion of decisions in favour of the gamma-Gompertz
  model. Left: Scenario $S_1$ for sample sizes $n_{90+}=$ 10,000,
  $n_{90+}=$ 20,000 and $n_{90+}=$ 105,000 (left to right) based on
  $\text{FIC}_{\text{MAE}}$ with $\mu=[\ln{h(100)}]''$
  (black-solid-circle), pre-test (red-dashed-cross) and
  $\text{AIC}^{\ast}$ (blue-dotted-triangle). Right: Scenarios $S_1$, $S_2$ and $S_3$ (left to right) all with $n_{90+}=$ 20,000 based on $\text{FIC}_{\text{MAE}}$ with $\mu=[\ln{h(100)}]''$ (black circle), $\mu=[\ln{h(110)}]''$ (red cross) and $\mu=\sigma^2$ (blue triangle).}
\label{fig:performance1}
\end{figure}

The left panel of Figure~\ref{fig:performance1} compares the performance of
the three selection approaches in scenario~$S_1$ ($\sigma^2=0.0625$) across
the various sample sizes. The $\text{FIC}_{\text{MAE}}$ with focus parameter
$\mu=[\ln{h(100)}]''$ clearly outperforms the other two methods, as it detects mortality deceleration more often. The proportion of correct decisions in favour of the gamma-Gompertz model increases with the sample size for all three methods, and is close to one for the setting with a large sample size. However, for the setting with a small (medium) sample size, the proportion of correct decisions based on the $\text{FIC}_{\text{MAE}}$ is 82.6\% (37.1\%) higher than that based on the $\text{AIC}^{\ast}$.

The right panel of Figure~\ref{fig:performance1} illustrates the
performance of the $\text{FIC}_{\text{MAE}}$ depending on the
magnitude of the frailty variance, and on the choice of the
focus parameter in the medium sample size setting. We display the results
for the focus parameters $\mu=\sigma^2$, $\mu=[\ln{h(100)}]''$ and
$\mu=[\ln{h(110)}]''$. The ability of the method to detect deviations from
the Gompertz hazard naturally decreases when the frailty variance
decreases. For scenario $S_2$, in which the frailty variance is about half as
large as it is in scenario $S_1$, the proportion of correct decisions is
about 35\% smaller than it is for $S_1$. If the true model is the Gompertz
model $(S_3)$, then the proportion of decisions in favour of the gamma-Gompertz
model is about 25\% for the medium sample size. As the
$\text{FIC}_{\text{MAE}}$ performs equally well for all three focus
parameters, the age~$y$ at which $\mu=[\ln{h(y)}]''$ is evaluated
does not seem to matter. It also turns out that the focus parameters
$\mu=\sigma^2$ and $\mu=[\ln{h(y)}]''$ perform better than, for
instance, $\mu=\ln{h(y)}$ or $\mu=S(y)$; as is shown in Section S.5 of the
supplementary material. Although the focus age $y$ did not affect the
results in the simulation study, other aspects may render one choice
more reasonable than another. In the medium-sized scenario $S_1$, in which around 20,000 individuals reach age 90, more than a thousand will, on average, also reach age 100, but fewer than 10 will reach age 110. Consequently, a focus age of $y=100$ will probably produce more reliable results than a focus age of $y=110$.

\begin{figure}[ht]
\centering
\includegraphics[width=\textwidth]{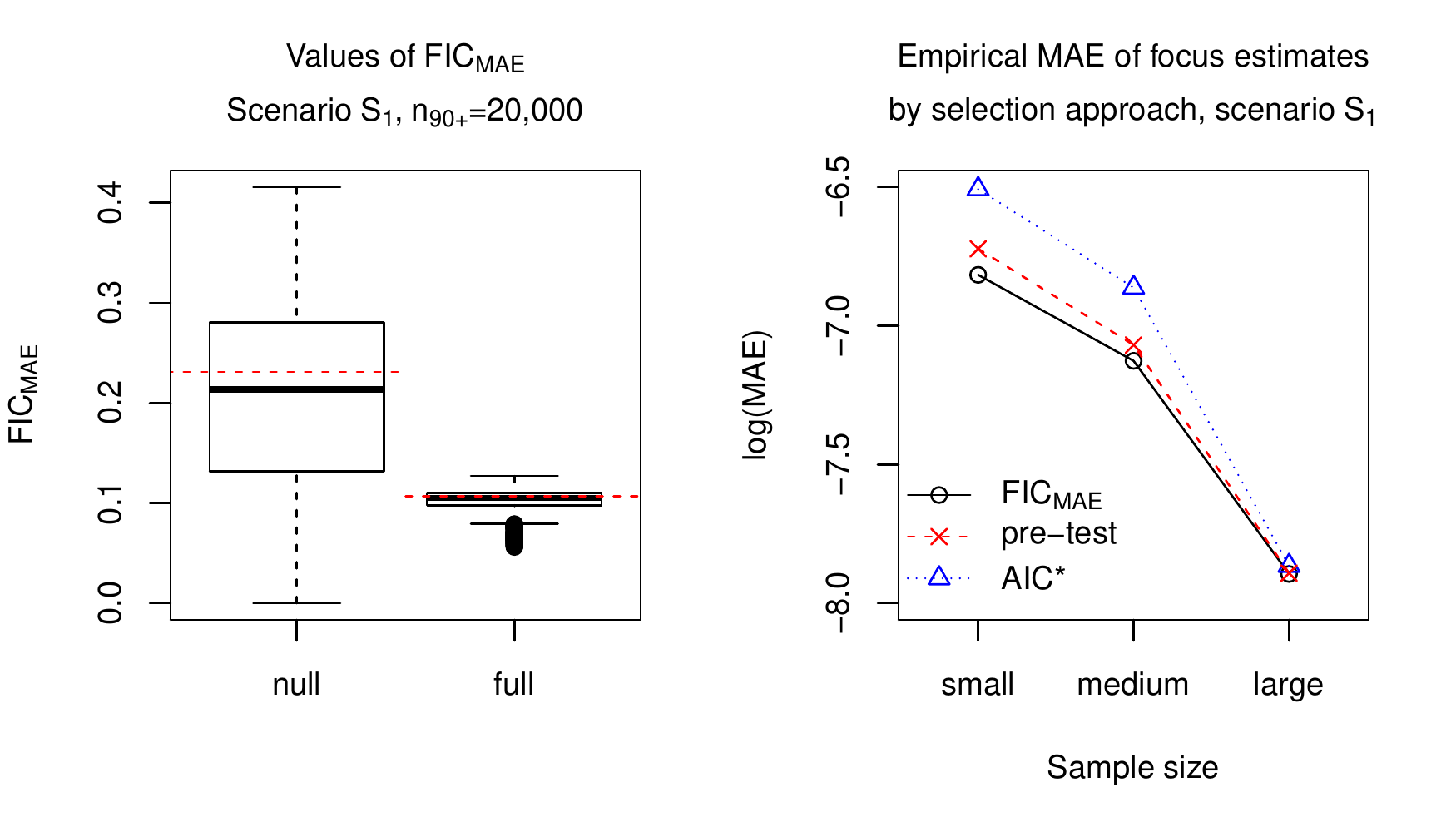}
\caption{Left: Box plots of $\text{FIC}_{\text{MAE}}$ values with $\mu=[\ln{h(100)}]''$ for the null and the full model in scenario~$S_1$ with $n_{90+}=$ 20,000 and empirical MAE of focus estimates~$\hat{\mu}$ (red-dashed). Right: Empirical MAE of selected $\hat{\mu}$ for $\mu=[\ln{h(100)}]''$ in scenario $S_1$ for sample sizes $n_{90+}=$ 10,000, $n_{90+}=$ 20,000 and $n_{90+}=$ 105,000 (left to right) based on $\text{FIC}_{\text{MAE}}$ (black-solid-circle), pre-test (red-dashed-cross), and $\text{AIC}^{\ast}$ (blue-dotted-triangle).}
\label{fig:performance2}
\end{figure}

The concept of the $\text{FIC}_{\text{MAE}}$ as an estimator of the limiting MAE of $\hat{\mu}$ is illustrated in the left panel of Figure~\ref{fig:performance2}, which shows a box plot of the $\text{FIC}_{\text{MAE}}$ values with $\mu=[\ln{h(100)}]''$ for 1,000 replications of the medium-sized scenario~$S_1$. We see that for both the null and the full model, the empirical MAEs of the estimators~$\hat{\mu}_{\text{null}}$ and $\hat{\mu}_{\text{full}}$ are close to the average of the respective FIC scores. As a consequence, the empirical MAE of the selected estimators in the 1,000 replications -- that is, $\hat{\mu}_{\text{full}}$ for those replications, where $\text{FIC}_{\text{MAE}}(\text{full})<\text{FIC}_{\text{MAE}}(\text{null})$ and $\hat{\mu}_{\text{null}}$ otherwise -- should be smaller than it is for other selection criteria. The right panel of Figure~\ref{fig:performance2} verifies for scenario~$S_1$, that the estimator $\hat{\mu}$ of $\mu=[\ln{h(100)}]''$ has the smallest empirical MAE when the model selection is based on the $\text{FIC}_{\text{MAE}}$ with $\mu=[\ln{h(100)}]''$, rather than on the pre-test or the $\text{AIC}^{\ast}$.

Overall, the findings of the simulation study support the claim that the proposed $\text{FIC}_{\text{MAE}}$ is a suitable tool for detecting mortality deceleration in the framework of the gamma-Gompertz model, which outperforms the competing approaches of the pre-test and $\text{AIC}^{\ast}$.

\section{Mortality of French-Canadians at high ages}\label{sec:appl}

As an application of the proposed methods, we analyse a highly reliable set of data on French-Canadians that was briefly introduced in Section~\ref{sec:Intro}. This data set is an expanded version of an earlier collection of verified mortality data on French-Canadian centenarians (see \citet{OuelletteBourbeau:2014} and \citet{Ouellette:2016} for further details) to which the deaths of individuals aged 90-99 were added. The data cover virtually all Catholic French-Canadians (20,917 females and 10,878 males) who were born in the Province of Quebec during the 1880-1896 period, and who died at ages 90 and above in Quebec between 1970 and 2009. These 1880-1896 birth cohorts were fully extinct by the end of 2009. The exact survival times in days were obtained by linking individual death certificates to corresponding birth registration documents taken from Quebec's parish register archives.

We fit the gamma-Gompertz model and the Gompertz model to the female
and the male data separately via maximum likelihood. For that purpose, we set
age 60 as the starting age of the models, and take into account left
truncation at age 90. Then, we choose between the gamma-Gompertz model
and the Gompertz model based on the $\text{AIC}^{\ast}$, pre-test and
$\text{FIC}_{\text{MAE}}$ for the focus parameters $\mu=\sigma^2$ and $\mu=[\ln{h(100)}]''$.

\begin{figure}[hbt]
\centering
\includegraphics[width=\textwidth]{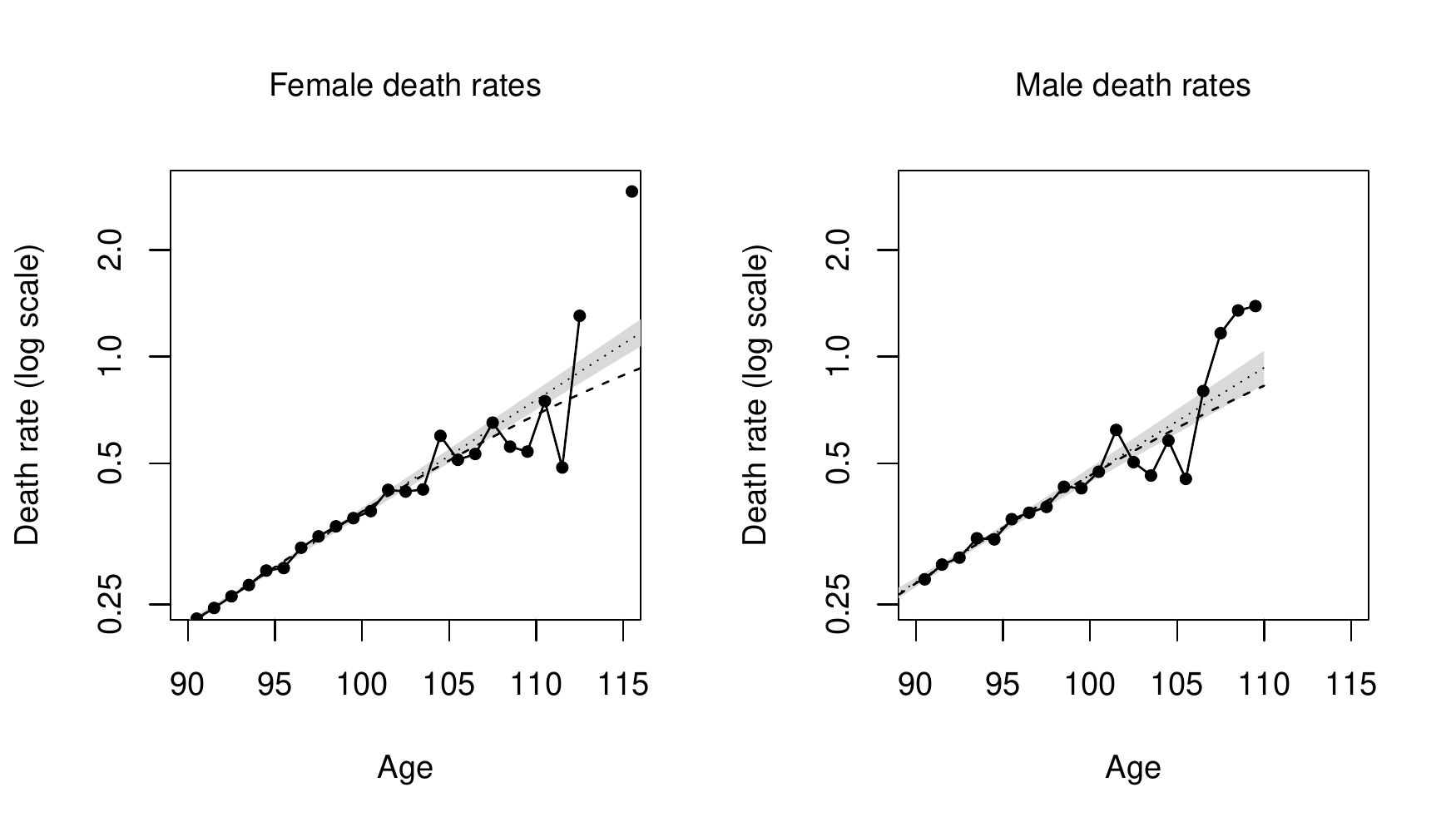}
\caption{Death rates (on log scale) of French-Canadian females (left) and males (right): observed death rates (solid-circle), gamma-Gompertz fit (dashed), Gompertz fit (dotted), and 95\%-confidence band for the Gompertz log-hazard (grey).}
\label{fig:Quebec}
\end{figure}

Figure~\ref{fig:Quebec} shows the fit of the gamma-Gompertz model and
the Gompertz model, respectively, to the empirical death rates (single
years of age) for the French-Canadian cohorts. The estimated frailty variance in the gamma-Gompertz model is $\hat{\sigma}^2=0.043$ for the female population and $\hat{\sigma}^2=0.037$ for the male population. A likelihood ratio test for $H_0\!: \sigma^2=0$ results in a $p$-value of $0.121$ for females and $0.283$ for males, such that the hypothesis of no mortality deceleration would not be rejected at the usual levels of significance.
Table~\ref{tab:MSV} also shows that based on the modified $\text{AIC}^{\ast}$, the Gompertz model is selected for both females and males. By contrast, based on the pre-test and the $\text{FIC}_{\text{MAE}}$, the gamma-Gompertz model is selected for females and the Gompertz model is selected for males. Hence, it appears that unlike other methods, the $\text{FIC}_{\text{MAE}}$ detects mortality deceleration in the female sample. Figure~\ref{fig:Quebec} also supports this finding of a deceleration in the female mortality rates.

\begin{table}[ht]
\centering
\caption{Values of different model selection criteria for the gamma-Gompertz model (GG) and the Gompertz model using data on French-Canadians.}
\begin{tabular}{lrrrr}
\toprule
 & \multicolumn{2}{c}{Females} & \multicolumn{2}{c}{Males}\\
 & \multicolumn{1}{c}{GG} & \multicolumn{1}{c}{Gompertz} & \multicolumn{1}{c}{GG} & \multicolumn{1}{c}{Gompertz}\\
\midrule
 $\text{AIC}^{\ast}$ & $101390.3$ & $\mathbf{101390.0}$ & $48364.10$ & $\mathbf{48363.01}$\\
 $\text{FIC}_{\text{MAE}}: \mu=\sigma^2$ & $\mathbf{4.065}$ & $6.200$ & $4.316$ & $\mathbf{3.890}$\\
 $\text{FIC}_{\text{MAE}}: \mu=[\ln{h(100)}]''$ & $\mathbf{0.098}$ & $0.149$ & $0.120$ & $\mathbf{0.108}$\\
\bottomrule
\end{tabular}
\label{tab:MSV}
\end{table}

\section{Discussion}\label{sec:disc}

Motivated by the issue of how mortality deceleration can be assessed at high ages,
we have extended the FIC, as introduced by \citet{ClaeskensHjort:2003},
to a non-standard setting in which we are choosing
between two models that differ by one parameter that takes a value on
the boundary of the parameter space if the smaller model is the true
model. We considered two versions of the FIC that aim to minimise the
limiting MAE or MSE of the estimator of the focus, respectively. When
targeting the MAE, we obtained the new model selection criterion
$\text{FIC}_{\text{MAE}}$. When targeting the MSE, the model selection
does not depend on the chosen focus, but a pre-test strategy was
defined. In addition, we presented the new $\text{AIC}^{\ast}$, which
reduces the bias of the original AIC that occurs when the selection
concerns a parameter that lies on the boundary of the parameter space
in the narrow model.

The proposed model selection criteria provide new tools for the assessment of mortality deceleration in the framework of the gamma-Gompertz model. While traditional approaches either have low power to detect mortality deceleration or are not valid in the presence of boundary-constrained parameters, the methods developed here are adapted to the non-standard setting. An advantage of the $\text{FIC}_{\text{MAE}}$ is that, by choosing an appropriate focus parameter, it can be targeted directly at the quantities that reveal mortality deceleration. We recommend using as the focus parameter the frailty variance or the second derivative of the log-hazard at some advanced age. Both potential choices readily translate into the presence or the absence of mortality deceleration, as the focus parameter takes a value of zero if there is no deceleration.

The results of our simulation studies indicate that the $\text{FIC}_{\text{MAE}}$, especially with the recommended choices of the focus parameter, outperforms the competing approaches of the pre-test and the $\text{AIC}^{\ast}$ in detecting mortality deceleration. This observation was made for different magnitudes of the frailty variance, and with different sample sizes. We found that the $\text{FIC}_{\text{MAE}}$ performs substantially better than the $\text{AIC}^{\ast}$, particularly for small samples. Moreover, in contrast to the other methods, the $\text{FIC}_{\text{MAE}}$ detected mortality deceleration in our sample of female French-Canadian Catholics born at the end of the $19^{\text{th}}$ century. It therefore appears that using the $\text{FIC}_{\text{MAE}}$ approach can bring new insights into the ongoing debate about mortality deceleration.

While the set-up in this article was restricted to individual-level
data, many studies on ageing rely on aggregated data in which death counts
and exposure times are available for given age-intervals. However, an
extension of the approach to aggregated data is straightforward if we
keep the assumption of the parametric model. Consequently, the new
tools for the assessment of mortality deceleration presented here will
be applicable to a variety of data sets collected for different human
and non-human populations. For data on humans, the application of the
Gompertz hazard is well-studied and well-established,
both across time and across populations.
For data on non-human species, we might want to consider relaxing the
assumption of a parametric model for the hazard.
More research is needed to understand how more flexible hazard shapes can be
incorporated by, for example, using splines and penalised likelihood.

Although our development of the $\text{FIC}_{\text{MAE}}$ was
motivated by the specific problem of assessing mortality deceleration,
the method could be used in a range of other contexts in which
there is a need to choose between parametric models that differ only by one
parameter with a boundary constraint, such as when assessing
heterogeneity in other proportional hazards frailty models, or when
choosing between a Poisson model and an over-dispersed negative binomial model. Linear mixed models are another model class where some parameters, in that case variance components, are restricted to be non-negative and where a focused search is useful \citep{Cunenetal:2019}.

\bibliography{FICstat}
\bibliographystyle{rss}

\end{document}


\begin{center}
\sffamily{\textbf{\LARGE{Web-based supporting materials for\\\vspace{0.5cm}
{Shifting attention to old age: Detecting mortality deceleration using focused model selection}}\\\vspace{0.5cm}
\Large{by Marie B\"ohnstedt, Hein Putter, Nadine Ouellette, Gerda Claeskens, and Jutta Gampe}}}\footnotetext[1]{\textit{Address for correspondence:} M.~B\"ohnstedt, Max Planck Institute for Demographic Research, Konrad-Zuse-Str.~1, 18057 Rostock, Germany\\\sffamily{E-mail: boehnstedt@demogr.mpg.de}}\\
\normalfont\normalsize
\vspace{0.5cm}
\noindent\textit{May 14, 2019}
\end{center}
\section{Computational issues}

The maximum likelihood estimates of the parameters of the
gamma-Gompertz model are determined via numerical optimisation in R
\citep{R}. Non-negativity of the parameter estimates is achieved by
maximising the log-likelihood over the log-transformed
parameters. Nevertheless, values of the frailty variance~$\sigma^2$
that are close to zero cause numerical difficulties. Here, we briefly
describe the steps that we took to increase the numerical stability
of the estimation problem.

We maximise the log-likelihood via the R-function \texttt{nlm()},
where we can also supply the analytic gradient of the objective
function. In addition, Taylor expansions of the log-likelihood and its
gradient are used if the current value of $\sigma^2$ is smaller than
$10^{-5}$. The numerically identified maximum~$\hat{\sigma}^2$ might
still depend on the starting value that was provided to the
optimisation routine. We therefore recommend running the optimisation with a number of different starting values for the frailty variance and choosing the fit with the largest value of the log-likelihood as the final estimate.

For calculating the $\text{FIC}_{\text{MAE}}$ values, we need estimates not only of the model parameters, but of the information matrix $J_{\text{full}}$. For that purpose, we derive analytically the matrix~$H(a,b,\sigma^2)$ of second-order derivatives of the log-likelihood for the gamma-Gompertz model, and again use a Taylor expansion if $\sigma^2<10^{-5}$. $J_{\text{full}}$ is then estimated as $-n^{-1}H(\hat{a},\hat{b},\hat{\sigma}^2)$; and $\hat{\kappa}^2$ is the bottom right element of its inverse.

\newpage
\section{Power of the likelihood ratio test}

A likelihood ratio test (LRT) for homogeneity in the gamma-Gompertz
model, where $H_0\!: \sigma^2=0$ and $H_1\!: \sigma^2>0$, may have
low power to detect mortality deceleration.
To illustrate this property, we summarise the results for two of the
scenarios that were  described in Section~4 of the main paper. In
particular, we show the extent to which a smaller underlying frailty
variance or a smaller sample size can decrease the power of the test,
which is performed at a significance level of 5\%.
We also compare the power of the LRT in a situation in which only
individuals who survived beyond age 90 can
be studied to a situation in which observations for individuals who
survived beyond age 80 or 85 are available.

\begin{figure}[hb]
\centering
\vspace{-0.25cm}
\includegraphics[width=\textwidth]{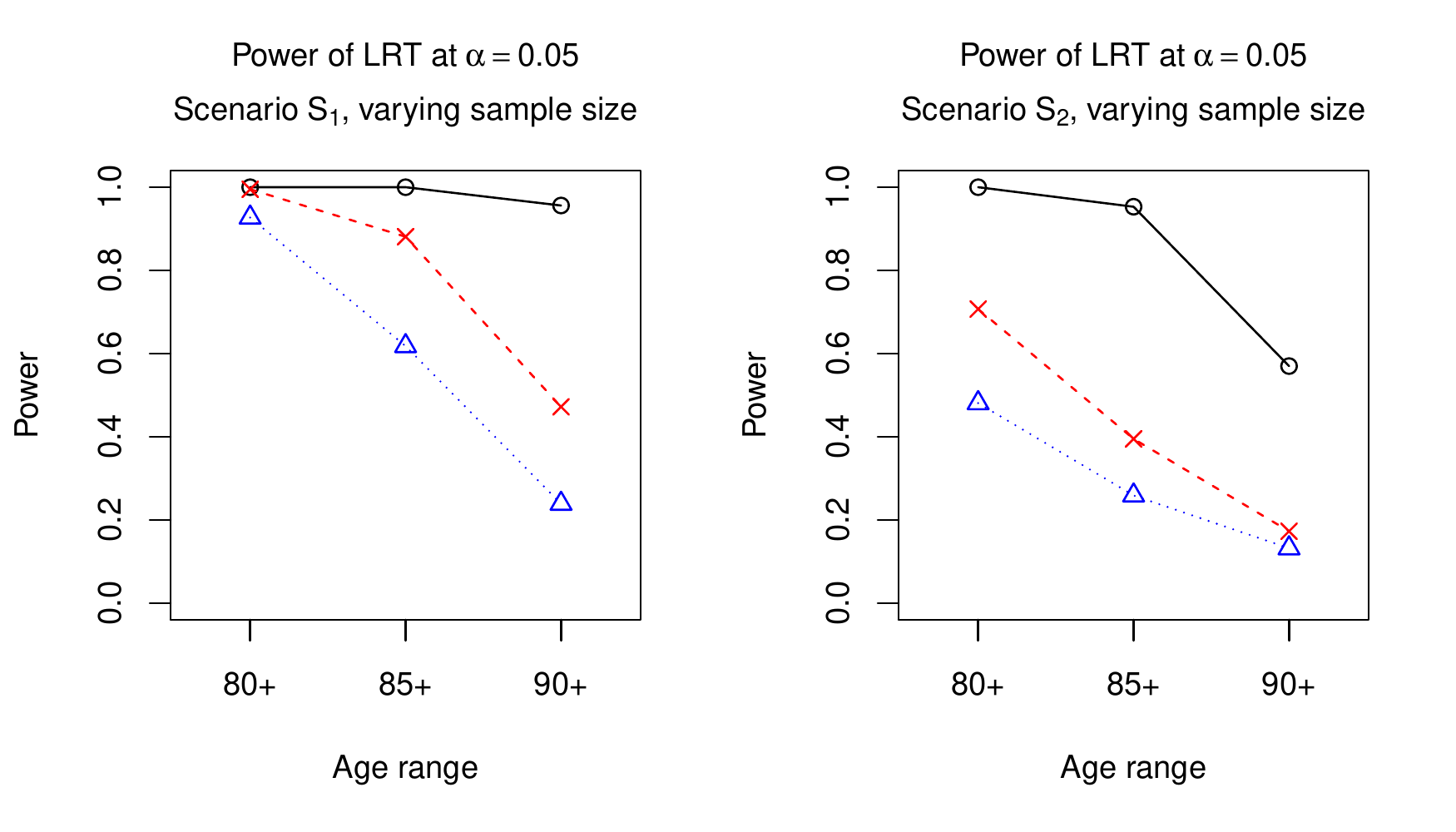}
\caption{Power of the LRT at the 5\% level to detect mortality deceleration in the gamma-Gompertz model depending on the age range of the data (left to right: 80+, 85+, or 90+). The depicted scenarios are $S_1$ (left) and $S_2$ (right) with the sample sizes $n_{90+}=$ 10,000 (blue-dotted-triangle), $n_{90+}=$ 20,000 (red-dashed-cross), and $n_{90+}=$ 105,000 (black-solid-circle).}
\label{fig:LRT}
\end{figure}

Figure~\ref{fig:LRT} illustrates how strongly the power
of the test is affected by the three features. The left panel displays
the results for scenario $S_1$ (frailty variance $\sigma^2=0.0625$),
while the right panel shows the results for scenario $S_2$ in which
the frailty variance was roughly halved ($\sigma^2=0.03)$.
Within each panel, we can see the loss
in power that occurs if only individuals who survived beyond age 90 (90+)
can be studied, instead of individuals who survived beyond age 80 (80+) or 85 (85+).
For example, in the medium-sized scenario $S_1$,
the power of the LRT decreases by more than 45\%
if the test is based on the 90+ subset instead of on the 85+ subset.

\newpage
\section{Local power of the LRT and the pre-test}

The local power of the LRT for homogeneity in the gamma-Gompertz model is derived in \citet{Boehnstedt:2018}. Under the sequence of local alternatives (see eq.~(2) in the main paper), the power of the LRT for $H_0\!: \sigma^2=0$ at level $\alpha$ based on a gamma-Gompertz sample of size~$n$ can be approximated by
\begin{equation}\label{eq:LRTpower}
1-\Phi\left(\Phi^{-1}(1-\alpha)-\frac{\delta}{\kappa}\right)=1-\Phi\left(\Phi^{-1}(1-\alpha)-\frac{\sqrt{n}\sigma^2}{\kappa}\right).
\end{equation}
The pre-test derived in Section~3.2 of the main paper selects the gamma-Gompertz model if $\hat{\delta}/\hat{\kappa}> 0.8399$. Due to $\hat{\delta}/\hat{\kappa}\stackrel{d}{\longrightarrow} \max{\left(0,D/\kappa\right)}$, we have $P\left[\hat{\delta}/\hat{\kappa}\leq z\right]\approx \Phi\left(z-\delta/\kappa\right)\mathbbm{1}_{\{z\geq0\}}$.
As a consequence, the power of the pre-test with critical region $\hat{\delta}/\hat{\kappa}>0.8399$ is determined as
\begin{equation}\label{eq:MSEpower}
P\left[\frac{\hat{\delta}}{\hat{\kappa}}>0.8399\Big|\text{fixed}~\delta\right]\approx  1-\Phi\left(0.8399-\frac{\delta}{\kappa}\right).
\end{equation}
Comparing~\eqref{eq:MSEpower} and~\eqref{eq:LRTpower}, we find that for large samples, the pre-test has approximately the same power as the LRT for $H_0\!: \sigma^2=0$ at level~$\tilde{\alpha}$ satisfying $\Phi^{-1}(1-\tilde{\alpha})=0.8399$, which is $\tilde{\alpha}=1-\Phi(0.8399)\approx 0.2005$.

\section{Derivation of $\text{FIC}_{\text{MAE}}$}

The $\text{FIC}_{\text{MAE}}$ of a model with focus estimator~$\hat{\mu}$, where $\sqrt{n}(\hat{\mu}-\mu_{\text{true}})\stackrel{d}{\longrightarrow}\Lambda$, is derived as an estimate of $\mathbb{E}[|\Lambda|]$.
For the null model, $\hat{\mu}_{\text{null}}$ converges to a normal distribution, $\Lambda_{\text{null}}=(\Lambda_0+\omega\delta)\sim\mathcal{N}(\omega\delta,\tau_0^2)$. Therefore, $\mathbb{E}[|\Lambda_{\text{null}}|]$ is calculated as the expected value of the folded normal random variable $|\Lambda_0+\omega\delta|$; that is,
\begin{equation}\label{eqA:MAEnull}
\mathbb{E}[|\Lambda_{\text{null}}|]=\mathbb{E}[|\Lambda_0+\omega\delta|]=2\tau_0\phi\left(\frac{\omega\delta}{\tau_0}\right) +2\omega\delta \left\{\Phi\left(\frac{\omega\delta}{\tau_0}\right)-\frac{1}{2}\right\}.
\end{equation}

For the full model, we have $\Lambda_{\text{full}}=\Lambda_0-\omega(D-\delta)\mathbbm{1}_{\{D>0\}}+\omega\delta\mathbbm{1}_{\{D\leq 0\}}$, with $D\sim\mathcal{N}(\delta,\kappa^2)$ independent of $\Lambda_0$, such that
\begin{align}\label{eqA:MAEfull}
\mathbb{E}[|\Lambda_{\text{full}}|]&=\mathbb{E}[|\Lambda_{\text{full}}|\mid D\leq 0] P[D\leq 0]+\mathbb{E}[|\Lambda_{\text{full}}|\mid D>0] P[D>0]\notag\\
&=\mathbb{E}[|\Lambda_0+\omega\delta|] \Phi\left(-\frac{\delta}{\kappa}\right)+\mathbb{E}[|\Lambda_0-\omega(D-\delta)|\mid D>0] \Phi\left(\frac{\delta}{\kappa}\right).
\end{align}
The first expectation is the same as~\eqref{eqA:MAEnull}. For
the computation of the second expectation, we define the normally
distributed random vector~$\boldsymbol{X}=(\Lambda_0,D)^T$ and its
affine transformation~$\boldsymbol{Y}=(\Lambda_0-\omega
(D-\delta),D)^T$, which is also normally distributed, with mean vectors~$\boldsymbol{\mu}_X=\boldsymbol{\mu}_Y=(0,\delta)^T$ and covariance matrices
\[\text{Cov}[\boldsymbol{X}]=\begin{pmatrix}\tau_0^2 & 0\\0 &\kappa^2\end{pmatrix}\quad\text{and}\quad\text{Cov}[\boldsymbol{Y}]=\begin{pmatrix}\tau_0^2+\omega^2\kappa^2 & -\omega\kappa^2\\-\omega\kappa^2 &\kappa^2\end{pmatrix}.\]
Then, $\mathbb{E}[|\Lambda_0-\omega(D-\delta)|\mid D>0]$ can be rewritten as
\begin{align}\label{eqA:absTrunc}
\mathbb{E}[|Y_1| \mid Y_2>0]&=\mathbb{E}[Y_1 \mid Y_1>0, Y_2>0]\frac{P[Y_1>0,Y_2>0]}{P[Y_2>0]}\notag\\&\hspace{3cm}+\mathbb{E}[-Y_1 \mid -Y_1\geq 0, Y_2>0]\frac{P[Y_1\leq 0,Y_2>0]}{P[Y_2>0]}.
\end{align}
The expected values of one component of a bivariate truncated normal distribution are more easily found for bivariate normal distributions with zero mean vectors, unit variances, and possible correlations. Transforming $\boldsymbol{Y}$ into such a normally distributed random vector $\boldsymbol{Z}=((\tau_0^2+\omega^2\kappa^2)^{-1/2} Y_1,(Y_2-\delta)/\kappa)^T$ with covariances $-\omega\kappa/\sqrt{\tau_0^2+\omega^2\kappa^2}$, and noting that
\[\mathbb{E}[Y_1 \mid Y_1>0, Y_2>0]=\sqrt{\tau_0^2+\omega^2\kappa^2} ~\mathbb{E}\left[Z_1 \Big| Z_1>0, Z_2>-\frac{\delta}{\kappa}\right],\]
we can apply the results of \citet{Tallis:1961} to obtain $\mathbb{E}[Y_1 \mid Y_1>0, Y_2>0]$ in~\eqref{eqA:absTrunc} as
\[\frac{\sqrt{\tau_0^2+\omega^2\kappa^2}}{P\left[Y_1>0,Y_2>0\right]} \left\{\frac{1}{\sqrt{2\pi}}\Phi\left(\frac{\delta}{\kappa}\cdot \frac{\sqrt{\tau_0^2+\omega^2\kappa^2}}{\tau_0}\right)-\frac{\omega\kappa}{\sqrt{\tau_0^2+\omega^2\kappa^2}}\,\phi\left(\frac{\delta}{\kappa}\right)\Phi\left(\frac{\omega\delta}{\tau_0}\right)\right\}.\]
Analogously, $\mathbb{E}[-Y_1 \mid -Y_1\geq 0, Y_2>0]$ in~\eqref{eqA:absTrunc} is computed as
\[\frac{\sqrt{\tau_0^2+\omega^2\kappa^2}}{P\left[Y_1\leq 0,Y_2>0\right]} \left\{\frac{1}{\sqrt{2\pi}}\Phi\left(\frac{\delta}{\kappa}\cdot \frac{\sqrt{\tau_0^2+\omega^2\kappa^2}}{\tau_0}\right)+\frac{\omega\kappa}{\sqrt{\tau_0^2+\omega^2\kappa^2}}\,\phi\left(\frac{\delta}{\kappa}\right)\Phi\left(-\frac{\omega\delta}{\tau_0}\right)\right\}.\]
Combining these two results, we find that $\mathbb{E}[|\Lambda_0-\omega(D-\delta)|\mid D>0]$ in \eqref{eqA:MAEfull} is equal to
\begin{equation}\label{eqA:MAEfull2}
\frac{\sqrt{\tau_0^2+\omega^2\kappa^2}}{\Phi\left(\frac{\delta}{\kappa}\right)}\cdot \sqrt{\frac{2}{\pi}}\Phi\left(\frac{\delta}{\kappa}\cdot \frac{\sqrt{\tau_0^2+\omega^2\kappa^2}}{\tau_0}\right)-\frac{\omega\kappa}{\Phi\left(\frac{\delta}{\kappa}\right)}\,\phi\left(\frac{\delta}{\kappa}\right) \cdot 2 \left\{\Phi\left(\frac{\omega\delta}{\tau_0}\right)-\frac{1}{2}\right\}.
\end{equation}
Inserting~\eqref{eqA:MAEnull} and \eqref{eqA:MAEfull2} into~\eqref{eqA:MAEfull} yields the postulated result
\begin{align*}
\mathbb{E}[|\Lambda_{\text{full}}|]&=\left[2\tau_0\phi\left(\frac{\omega\delta}{\tau_0}\right) +2\omega\delta \left\{\Phi\left(\frac{\omega\delta}{\tau_0}\right)-\frac{1}{2}\right\}\right]\left\{1-\Phi\left(\frac{\delta}{\kappa}\right)\right\}\\
& \hspace{0.5cm}+\sqrt{\tau_0^2+\omega^2\kappa^2}\cdot \sqrt{\frac{2}{\pi}}\Phi\left(\frac{\delta}{\kappa}\cdot \frac{\sqrt{\tau_0^2+\omega^2\kappa^2}}{\tau_0}\right)-\omega\kappa\,\phi\left(\frac{\delta}{\kappa}\right) \cdot 2 \left\{\Phi\left(\frac{\omega\delta}{\tau_0}\right)-\frac{1}{2}\right\}.
\end{align*}

\section{Other focus parameters}

\begin{figure}[hb]
\centering
\includegraphics[width=\textwidth]{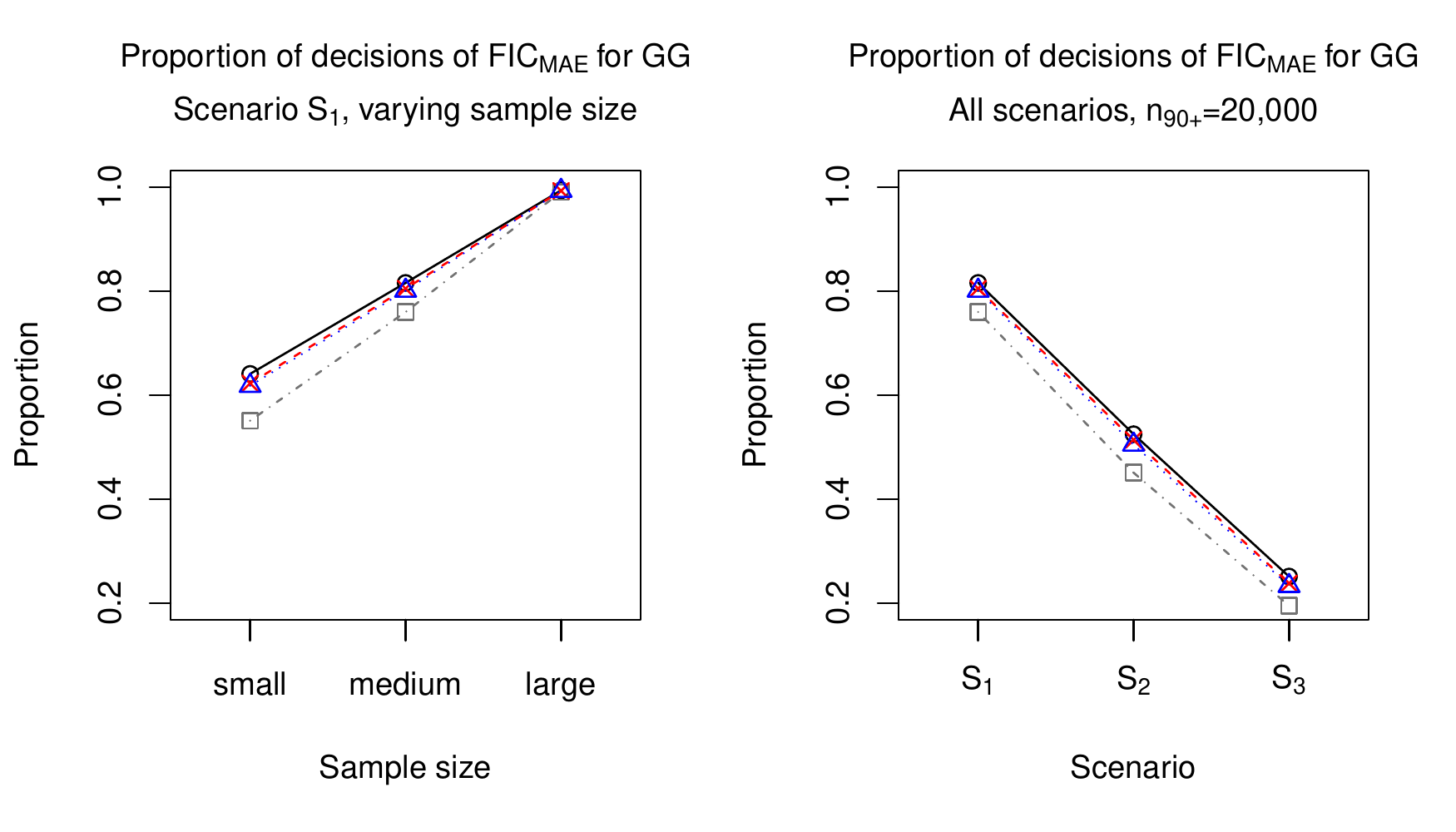}
\caption{Proportion of decisions in favour of the gamma-Gompertz model based on $\text{FIC}_{\text{MAE}}$ with $\mu=[\ln{h(100)}]''$ (black-solid-circle), $\mu=\ln{h(100)}$ (grey-dot-dashed-square), $\mu=\ln{h(110)}$ (red-dashed-cross) and $\mu=S(100)$ (blue-dotted-triangle). Left: Decisions in scenario $S_1$ for sample sizes $n_{90+}=$ 10,000, $n_{90+}=$ 20,000 and $n_{90+}=$ 105,000 (left to right). Right: Decisions in scenarios $S_1$, $S_2$ and $S_3$ (left to right) all with $n_{90+}=$ 20,000.}
\label{fig:foci}
\end{figure}

In our simulation studies, we also assessed the performance of the $\text{FIC}_{\text{MAE}}$ for several other focus parameters, such as quantiles of the survival distribution or the log-hazard and the survival function at different advanced ages. Overall, the frailty variance, $\mu=\sigma^2$, and the second derivative of the log-hazard, $\mu=[\ln{h(y)}]''$, yielded the best results. Figure~\ref{fig:foci} illustrates the proportion of decisions in favour of the gamma-Gompertz model in several settings, when the focus is placed on the second derivative of the log-hazard at age 100, the log-hazard at age 100 or 110, or the survival function at age 100. While the choice of $\mu=[\ln{h(100)}]''$ again results in the highest proportion of decisions in favour of the gamma-Gompertz model, the choice of $\mu=S(100)$ performs almost as well. When the focus is put on the log-hazard, the age at which the function is evaluated apparently makes a difference, in that $\mu=\ln{h(110)}$ leads to a better performance of the $\text{FIC}_{\text{MAE}}$ than $\mu=\ln{h(100)}$. However, survival beyond age 110 is relatively rare in some of our simulated settings, and we should be careful when putting the focus on ages for which there are too few data points.

\section{Impact of the age range on the performance of the $\text{FIC}_{\text{MAE}}$}

\begin{figure}[hb]
\centering
\includegraphics[width=\textwidth]{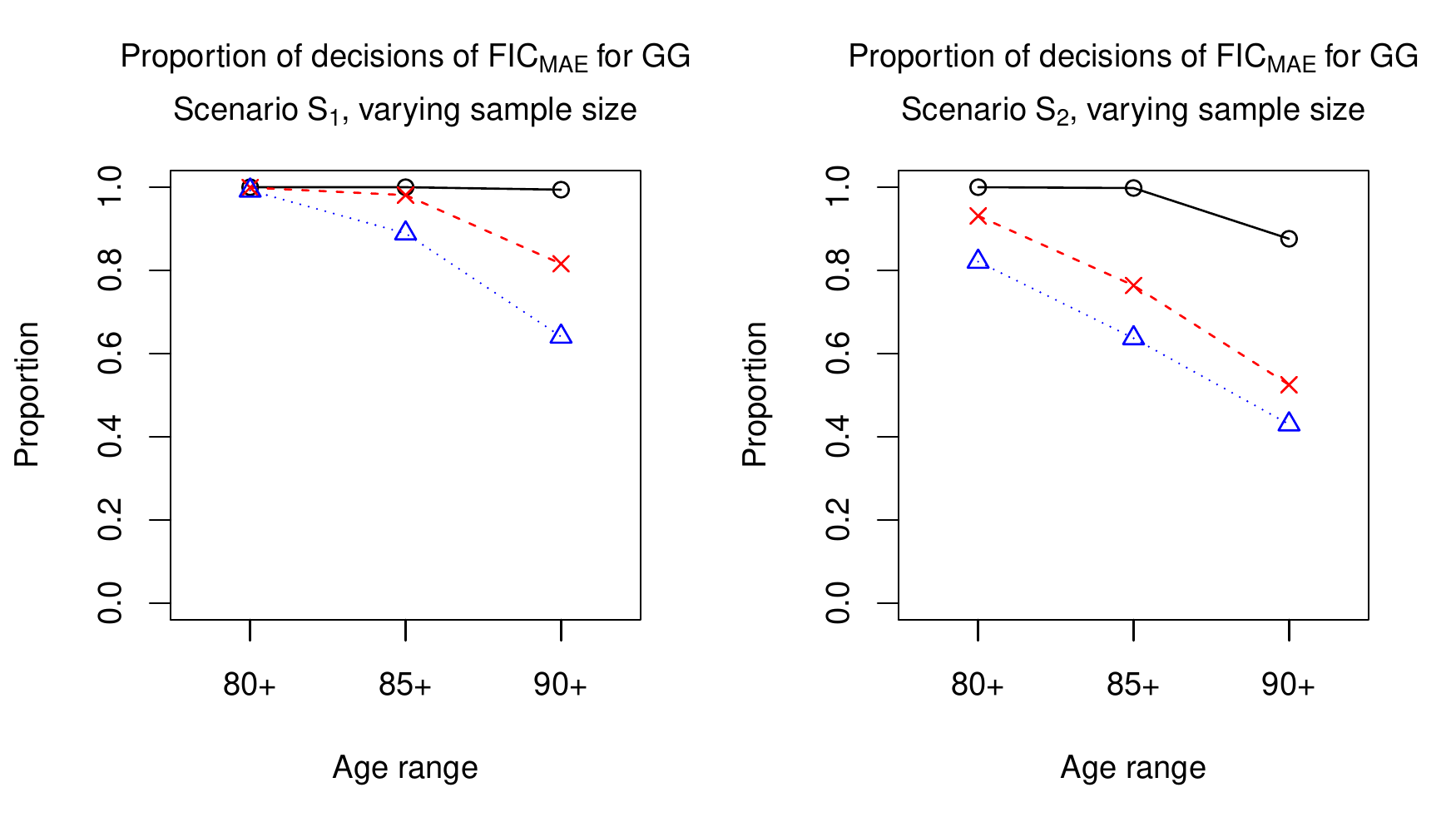}
\caption{Proportion of correct decisions in favour of the gamma-Gompertz model based on the $\text{FIC}_{\text{MAE}}$ with $\mu=[\ln{h(100)}]''$ depending on the age range of the data (left to right: 80+, 85+, or 90+). The depicted scenarios are $S_1$ (left) and $S_2$ (right) with sample sizes $n_{90+}=$ 10,000 (blue-dotted-triangle), $n_{90+}=$ 20,000 (red-dashed-cross) and $n_{90+}=$ 105,000 (black-solid-circle).}
\label{fig:ageRange}
\end{figure}

In the main paper, we have, motivated by the data application, considered only samples of individuals who survived beyond age 90. However, the amount of heterogeneity in mortality risk within the population decreases with age due to selection. Therefore, it is of interest to study the performance of the $\text{FIC}_{\text{MAE}}$ according to the age range of the sample. Figure~\ref{fig:ageRange} depicts the proportion of correct decisions in favour of the gamma-Gompertz model based on the $\text{FIC}_{\text{MAE}}$ with $\mu=[\ln{h(100)}]''$ in different settings when the sample consisted of all individuals who had reached at least age 80, 85, or 90. We see that, in general, the probability of detecting mortality deceleration increases if the sample covers a wider age range. For scenario~$S_1$ with the target sample size of $n_{90+}=$ 10,000, the proportion of correct decisions increases by more than a third if we observe all individuals who had reached at least age 85 instead of only those individuals who had reached at least age 90. Both the larger sample size of the 85+ subset and the greater amount of heterogeneity in the mortality risk of this subset played a part in this result.

\bibliography{FICstatSuppl}
\bibliographystyle{rss}